\documentclass[preprints,article,accept,pdftex,moreauthors]{Definitions/mdpi}
\newcommand{\summ}{\sum_{i=1}^{W}}
\newcommand{\dd}{\textrm{\,d}}

\firstpage{1}
\pubvolume{1}
\issuenum{1}
\articlenumber{0}
\pubyear{2023}
\copyrightyear{2023}
\externaleditor{ }
\datereceived{ }
\daterevised{ } % Comment out if no revised date
\dateaccepted{ }
\datepublished{ }
\hreflink{https://doi.org/} % If needed use \linebreak
\Title{A Note on the Connection between Non-Additive Entropy and~$h$-Derivative}

\TitleCitation{A Note on the Connection between Non-Additive Entropy and $h$-Derivative}

\Author{Jin-Wen Kang$^{1}$\orcidB{}, Ke-Ming Shen$^{1,2,}$*\orcidA{} and Ben-Wei Zhang $^{1,}$*\orcidC{}}

\AuthorNames{Jin-Wen Kang, Ke-Ming Shen and Ben-Wei Zhang}

\AuthorCitation{Kang, J.W.; Shen, K.M.; Zhang, B.W.}
\address{%
$^{1}$ \quad Key Laboratory of Quark \& Lepton Physics (MOE), Institute of Particle Physics,
 Central China Normal University, Wuhan 430079, China; kangjinwen@mails.ccnu.edu.cn\\
$^{2}$ \quad School of Science, East China University of Technology, Nanchang 330013, China}

\corres{Correspondence: shen$\_$keming.ecut@hotmail.com~(K.-M.S.); bwzhang@mail.ccnu.edu.cn~(B.-W.Z.)}
\abstract{In order to study as a whole a wide part of entropy measures, we introduce a two-parameter non-extensive entropic form with
respect to the $h$-derivative, which generalizes the conventional Newton--Leibniz calculus.
This new entropy, $S_{h,h'}$, is proved to describe the non-extensive systems and recover several types of well-known
non-extensive entropic expressions, such as the Tsallis entropy, the Abe entropy, the Shafee entropy, the Kaniadakis entropy and even the
classical Boltzmann--Gibbs one.
As a generalized entropy, its corresponding properties are also analyzed.
}


\begin{document}

%%%%%%%%%%%%%%%%%%%%%%%%%%%%%%%%%%%%%%%%%%
%\setcounter{section}{-1} %% Remove this when starting to work on the template.
\section{Introduction}

Since it was proposed over one hundred years ago, the conventional Boltzmann--Gibbs (BG) statistics has been developed very delicately and successfully with wide applications in many disciplines.
During the last few decades, however, people noticed that more and more systems are difficult to be described by this simple BG distribution, such as the long-range interactions~\cite{Dauxois2002}, the gravitational systems~\cite{salzberg1965exact}, the L\'evy flights and fractals~\cite{montroll1983maximum}, and so on~\cite{TSALLIS1995539}.
In order to cope with this challenge, some attempts have been made to generalize the BG statistics.
Among them, the most investigated formalism is the non-extensive entropy.
It was inspired by the geometrical theory of multi-fractals and its systematic use of powers of probabilities by  {C. Tsallis}
~\cite{Tsallis:1987eu}:
\begin{eqnarray}
	S_q= k_B\frac{1-\summ p_i^q}{q-1},
\label{equ:tsallisdefine}
\end{eqnarray}
where $k_B$ is the Boltzmann constant (hereafter we assume $k_B=1$ for simplicity) and $q$ stands for the Tsallis non-extensive parameter.
It describes the departure of non-extensive statistics from the BG one.
This entropy goes back to the usual BG form when $q\to 1$.
For more than two decades of researches and developments, the Tsallis entropy has been successfully applied to various domains:
physics, chemistry, economics, computer science, biosciences, linguistics, and so
on~\cite{tsallis2009,PLASTINO1995347,Tsallis:1987eu,Biro:2014ata,Shen:2017qwo,Shen:2017etj}.
%Some concrete and simple examples are as follows. %
For the average charged-hadron yields in inelastic non-single-diffractive events, V. Khachatryan et al. observe it as the Tsallis distribution~\cite{CMS:2010tjh, CMS:2010wcx}
\begin{eqnarray}
E\frac{{\rm d}^3N_{ch}}{{\rm d}p^3}=\frac{1}{2\pi p_T}\frac{{\rm d}^2N_{ch}}{{\rm d}\eta\,{\rm d}p_T}=C\frac{{\rm d}N_{ch}}{{\rm d}y}\left(
	1+\frac{E_T}{nT}
\right)^{-n},
\label{spectra}
\end{eqnarray}
where $E\frac{{\rm d}^3N_{ch}}{{\rm d}p^3}$ is for the function of spectra with $E$ for the total energy of the particle and $p$ for its momentum, $\eta$ denotes the pseudorapidity with $y$ for the rapidity, $p_T$ stands for the transverse momentum, $C$ is for its normalization constant, $T$, a variational parameter representing the temperature when the system reaches equilibrium,
$n$ is the fitting parameter which connects with Tsallis' $q$ by $n=1/(1-q)$, $y=0.5\ln[(E+p_z)/(E-p_z)], E_T=\sqrt{m^2+p_T^2}-m$, and $m$ is the charged pion mass. The data fitting results show that the
Tsallis distribution can well-describe both the low-$p_T$ exponential and the high-$p_T$ power-law behaviors~\cite{CMS:2010tjh, CMS:2010wcx}.
One application in astrophysics is the study of the distribution of asteroid rotation periods from different regions of
the solar system and diameter distributions of near-Earth asteroids (NEAs)~\cite{Betzl:2012}. A. S. Betzler and E. P. Borges analyze two samples from different years. They discover that the distribution of diameters of NEAs obeys a Tsallis-like distribution, and the rotation periods of asteroids can be well-approximated by a Tsallis--Gaussian function. According to the first conclusion, there should be
\mbox{$994\pm30$ NEAs} with diameters greater than $1~{\rm km}$~\cite{Betzl:2012}. In another example, Y. Wang and J. Du study the viscosity of light
charged particles in weakly
ionized plasma with the power-law Tsallis-distributions using the generalized Boltzmann equation of transport and the motion equation of hydrodynamics~\cite{wang2018viscosity}.

\textls[-10]{The Tsallis entropy is indeed not unique.
By now, a lot of different expressions of the non-additive entropies have been proposed, for instance, the Kaniadakis entropy~\cite{kaniadakis2001non}, the Shafee entropy~\cite{shafee2004generalized}, the $q - q^{-1}$ symmetric modification of Tsallis entropy~\cite{abe1997note}, and the two-parameter $(q,q')$-entropy~\cite{borges1998family}.
%For more details of the definitions of these entropies please see the Appendix.
These expressions were obtained in quite different ways and investigated by distinct motivations.
Therefore, it will be of great interest to find the relationship among these formulas or to find a simple formula to study them as a whole.}

In this paper, we first introduce a two-parameter non-additive entropy, $S_{h,h'}$, based on the $h$-derivative.
The $h$-derivative is known as a mathematical generalization of the normal Newton--Leibniz calculus.
We address that $S_{h,h'}$ unifies different types of expressions of non-extensive entropies; namely, it can connect a family of non-extensive entropies.
On the other hand, we also discuss its properties in order to better understand this newly established non-additive entropic function.

\section{\label{sec:hderivative}\boldmath{$h$}-Derivative}
In the conventional mathematical theory, the Newton--Leibniz derivative is defined as:
\begin{equation}
\label{equ:newton-derivative}
Df(x)\equiv\frac{{\rm d}f(x)}{{\rm d}x}=\lim_{\delta\rightarrow
0}\frac{f(x+\delta)-f(x)}{\delta}.
\end{equation}
Classically, most of the physical quantities are continuous, and it is natural to apply the Newton--Leibniz derivative.
In quantum physics, on the other hand, all the physical quantities will be quantized; people then try to develop quantum calculus, which utilizes the discrete forms of derivatives instead and presents a generalization of this Newton-Leibniz derivative.

One formalism of quantum calculus is the $h$-derivative~\cite{kac2002quantum}.
For an arbitrary function $f(x)$, its $h$-differential is defined as follows:
\begin{equation}
	\label{equ:h-differential}
	{\rm d}_hf(x)=f(x+h)-f(x).
\end{equation}
It is easily verified that
\begin{equation}
	\label{equ:dh-x}
	{\rm d}_hx=h,
\end{equation}
and
\begin{equation}
	{\rm d}_h\left(f(x)g(x)\right)=f(x+h){\rm d}_hg(x)+g(x){\rm d}_hf(x).
\end{equation}
Thus, can we obtain the $h$-derivative of $f(x)$:
\begin{equation}
	\label{equ:h-derivative}
	D_hf(x)\equiv\frac{{\rm d}_hf(x)}{{\rm d}_hx}=\frac{f(x+h)-f(x)}{h}.
\end{equation}
When $f(x)$ is differentiable, the following property is obviously obtained:
\begin{equation}
	\lim_{h\rightarrow 0}D_hf(x)=\frac{{\rm d}f(x)}{{\rm d}x},
\end{equation}
which is nothing but the definition of the conventional Newton--Leibniz derivative.
Note that we need the function $f(x)$ to be continuous for the Newton--Leibniz derivative,
but this requirement becomes unnecessary for the $h$-derivative.

Next, some basic rules of this $h$-derivative are listed:
%\paragraph{\hl{Sum and difference} %MDPI: \paragraph{ } is used for heading 4, it is below heading 1, it is incorrect, Please check if it should be formatted as subsection to be level 2.
    %Author: We have changed \paragraph{ } to enumerate environment.
%}
\begin{enumerate}
    \item Sum and difference\\
    Considering the sum and difference rules of the $h$-derivative, we have
    \begin{equation}
        D_h\left[f(x)\pm g(x)\right]= D_hf(x)\pm D_hg(x).
    \end{equation}
    \item Product and quotient rules\\
    As for the product and quotient rules,
    \begin{eqnarray}
        D_h\left[f(x)g(x)\right]&=&f(x)D_hg(x)+g(x+h)D_hf(x), \\
        D_h\left[\frac{f(x)}{g(x)}\right]&=&\frac{g(x)D_hf(x)-f(x)D_hg(x)}{g(x)g(x+h)}.
    \end{eqnarray}
    \item \textit{h}-derivative of elementary functions\\
    Some other basic calculations of it are expressed:
    \begin{equation}
        D_hC=0~~~~(\text{here $C$ is constant}),
    \end{equation}
    \begin{equation}
        D_h x = \frac{(x+h)-x}{h} = 1,
    \end{equation}
    \begin{equation}
        D_h x^n = \sum_{k=0}^{n-1}\frac{n!}{k!(n-k)!}x^k h^{n-k-1}~~~~(n\in\mathbb{N}),
    \end{equation}
    \begin{equation}
        D_h \frac{1}{x} = -\frac{1}{x^2+hx},
    \end{equation}
    \begin{equation}
        D_h\frac{1}{x^n} = \frac{1}{h(x+h)^n}-\frac{1}{hx^n},
    \end{equation}
    \begin{equation}
        D_h {\rm e}^{bx} = \frac{{\rm e}^{bh}-1}{h}{\rm e}^{bx}~~~~(\text{here~}
        b\in\mathbb{R}),
    \end{equation}
    \begin{equation}
        D_h a^{bx} = \frac{a^{bh}-1}{h}a^{bx}~~~~(\text{here~} b\in \mathbb{R}).
    \end{equation}
\end{enumerate}

%\paragraph{\hl{Product and quotient rules}}
%As for the product and quotient rules,
%\begin{eqnarray}
%	D_h\left[f(x)g(x)\right]&=&f(x)D_hg(x)+g(x+h)D_hf(x), \\
%	D_h\left[\frac{f(x)}{g(x)}\right]&=&\frac{g(x)D_hf(x)-f(x)D_hg(x)}{g(x)g(x+h)}.
%\end{eqnarray}

%\paragraph{\hl{\textit{h}-derivative of elementary functions}} Some other basic calculations of it are expressed:
%\begin{equation}
%	D_hC=0~~~~(\text{here $C$ is constant}),
%\end{equation}
%\begin{equation}
%	D_h x = \frac{(x+h)-x}{h} = 1,
%\end{equation}
%\begin{equation}
%	D_h x^n = \sum_{k=0}^{n-1}\frac{n!}{k!(n-k)!}x^k h^{n-k-1}~~~~(n\in\mathbb{N}),
%\end{equation}
%\begin{equation}
%	D_h \frac{1}{x} = -\frac{1}{x^2+hx},
%\end{equation}
%\begin{equation}
%	D_h\frac{1}{x^n} = \frac{1}{h(x+h)^n}-\frac{1}{hx^n},
%\end{equation}
%\begin{equation}
%	D_h {\rm e}^{bx} = \frac{{\rm e}^{bh}-1}{h}{\rm e}^{bx}~~~~(\text{here~}
%	b\in\mathbb{R}),
%\end{equation}
%\begin{equation}
%	D_h a^{bx} = \frac{a^{bh}-1}{h}a^{bx}~~~~(\text{here~} b\in \mathbb{R}).
%\end{equation}

\begin{figure}[H]
\includegraphics[width=0.9\linewidth]{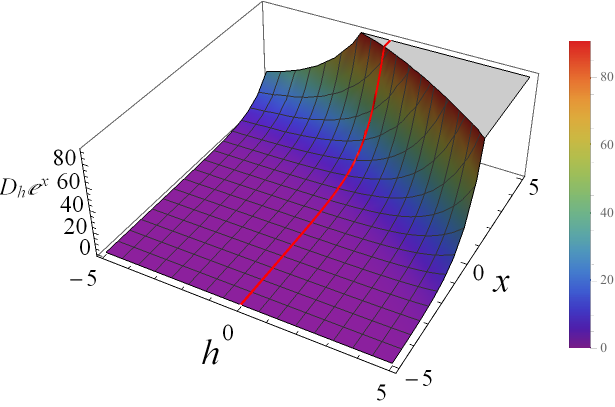}
\caption{The behavior of $D_h{\rm e}^x$ when $h$ varying from $-5$ to
$5$. The red line denotes $({\rm e}^x)'$.}
\label{fig:exp-h-derivative}
\end{figure}

In Figure~\ref{fig:exp-h-derivative}, we illustrate the behavior of $D_h{\rm e}^x$ at different values of $h$ as an example.
We could see that it behaves as an exponential when $h=0$.
For any fixed values of $h$, $D_h{\rm e}^x$ is a monotonically increasing function with respect to the variable $x$.
The values of this derivative also increase when the parameter $h$ becomes larger.

With the definition of $h$-derivative, V. Kac and P. Cheung~\cite{kac2002quantum} developed a type of quantum calculus, known as $h$-calculus.
As a matter of fact, an operator such as $h$-derivative is called the forward difference quotient operator.
Analogously, it also has the backward difference quotient operator $\nabla_h$ and the central difference quotient operator $\delta_h$, defined as~\cite{jagerman2000difference}
 \begin{equation}
	 \nabla_h f(x)=\frac{f(x)-f(x-h)}{h},
 \end{equation}
 \begin{equation}
	 \label{equ:central-diff}
	 \delta_h f(x)=\frac{f(x+\dfrac{1}{2}h)-f(x-\dfrac{1}{2}h)}{h}.
 \end{equation}
Note that the regular vector differential operator $\nabla$ has been generalized based on $h$-derivative.
We then explore the connection between the $h$-derivative entropy and its modified forms.

\section{\textit{h}-Derivative and Non-Additive Entropy}
%We then show there is a remarkable relation between \textit{h}-derivative and nonextensive entropy.
In order to generalize the non-additive entropic forms, one could utilize Equation~(\ref{equ:h-derivative}) and give out the following equation:
\begin{equation}
	S_h=-D_h\sum_{i=1}^{W}p_i^x\Bigg|_{x=1}=-\frac{\summ p_i^{1+h}-1}{h},
\end{equation}
with the normalization condition $\summ p_i =1$.
When $h\to 0$, it will go back to the usual BG entropy.
Note that it also recovers the Tsallis non-extensive entropy, $S_q$, cf. Equation (\ref{equ:tsallisdefine}) under the transformation of $h=q-1$.

Following the ways of the central difference quotient operator of Equation~(\ref{equ:central-diff}), we define a new form of two-parameter $(h,h')$-derivative,
\begin{equation}
	\label{equ:def-two-h-diff}
	D_{h,h'}f(x)=\frac{f(x+h)-f(x-h')}{h+h'}~~~~(h,h'\in\mathbb{R}).
\end{equation}
The corresponding $(h,h')$-entropy is
\begin{equation}
S_{h,h'}=-D_{h,h'}\sum_{i=1}^{W}p_i^x\Bigg|_{x=1}=-\sum_{i=1}^{W}
	\frac{p_i^{1+h}-p_i^{1-h'}}{h+h'}.
\end{equation}

Similarly, when $h=h'\rightarrow 0$, the entropy $S_{h,h'}$ returns to the BG one.
It is shown that,
\begin{eqnarray}
	\lim_{h=h'\rightarrow 0} S_{h,h'} &=& -\lim_{h=h'\rightarrow 0}\summ
		\frac{p_i^{1+h}-p_i^{1-h'}}{h+h'} \nonumber \\
		&=& -\lim_{h=h'\rightarrow 0}\summ p_i\frac{{\rm e}^{h\ln p_i}-{\rm e}^{-h'\ln p_i}}{h+h'} \nonumber \\
		&=& -\lim_{h=h'\rightarrow 0}\summ p_i\frac{{\rm e}^{h \ln p_i} \ln p_i+
					{\rm e}^{-h'\ln p_i}\ln p_i}{2} \nonumber \\
			&=& -\summ p_i\ln p_i = S_{BG}.
\end{eqnarray}
Note that L'Hospital's rule has been applied within the formula ${\rm d}e^{\alpha x}/{\rm d}x=\alpha e^{\alpha x}$ for the last step in the above.

It is constructive to explore the connections with the already known statistical distributions.
For example, the Tsallis entropy is obtained by $h=q-1, h'=0$.
While taking $h=q-1, h'=1-q^{-1}$, we can obtain the Abe entropy Equation~(\ref{equ:abeentropy}) (see the discussion in the Appendix \ref{app1})~\cite{abe1997note}.
The non-extensive entropy given in Equation~(\ref{equ:borgesentropy}) proposed by Borges and Roditi~\cite{borges1998family} (also see the Appendix \ref{app1})  can then be recovered with the relationship of $h=q-1, h'=1-q'$.
Although the non-extensive entropy of Borges and Roditi and our two-parameter $(h,h')$-entropy have similar forms, we gained them using different mathematical methods. Specifically, we used $(h,h')$-derivative developed by ourselves, which differs from the $q$-calculus used by Borges and Roditi. In addition to the difference in the form of expression between the two-type derivative, a conspicuous point is that our two-parameter $(h,h')$-derivative does not require the function $f(x)$ to be continuous and differentiable at $x=0$.

By assuming $h'=h$, we could also obtain another new form of entropy $S_{h,h}$,
% \begin{equation}
% 	\xcancel{S_{h,h}=-\summ \frac{p_i^{1+h}-p_i^{1-h}}{2h} \ ,}
% \end{equation}
which is obviously invariant under the interchange $h\leftrightarrow -h$.
As a matter of fact, it is nothing new but the well-known Kaniadakis non-extensive $\kappa$-entropy~\cite{kaniadakis2001non}. %\redsout{:}\add{.}
% \begin{equation}
% \xcancel{S_{\kappa}^{K}= -\sum_{i=1}^{W}p_i\ln_{\kappa}^{K}p_i \ ,}
% \end{equation}
% \redsout{where}
% \begin{equation}
% \xcancel{\ln_{\kappa}^{K}p_i\equiv\frac{p_i^{\kappa}-p_i^{-\kappa}}{2\kappa}
% ~~~~(\ln_0^Kp_i=\ln p_i) \ .}
% \end{equation}
Last but not least, it is set that $h'\rightarrow -h$ and $h'=-h+\delta$.
Considering the limit of $\delta \rightarrow 0$ and $h'\rightarrow -h$, we could also cover the exact Shafee entropy~\cite{shafee2004generalized,shafee2007lambert} by
taking the transformation of $q=h+1$.
% \begin{eqnarray}
% 	\xcancel{\lim_{h'\rightarrow -h}S_{h,h'}}&=& \xcancel{-\sum_{i=1}^{W}\lim_{\delta\rightarrow 0}
% 				\frac{p_i^{1+h}-p_i^{1+h-\delta}}{h-h+\delta} \nonumber }\\
% 	&=& \xcancel{ -\sum_{i=1}^{W}p_i\lim_{\delta\rightarrow 0}\frac{{\rm e}^{h\ln
% 	p_i}-{\rm e}^{(h-\delta)\ln p_i}}{\delta} \nonumber }\\
% 	&=& \xcancel{ -\sum_{i=1}^{W}p_i^{1+h}\ln p_i \ . }
% \end{eqnarray}

In Table~\ref{tab:table1}, we summarize different entropy functions, which can be represented by this two-parameter $S_{h,h'}$ entropy through taking different values of $h$ and $h'$.
In addition, by choosing $h'=-1/h$ the function $S_{h,h'}$ becomes
\begin{equation}
S_{h,-1/h} = -\sum_{i=1}^{W} \frac{p_i^{1+h} - p_i^{1-1/h}}{h+1/h},
\end{equation}
Note that this entropic form looks much similar to Abe entropy~\cite{abe1997note}, but it is totally different in fact that Abe entropy cannot be recovered only by exchanging $q$ and $h$ when comparing them.
Hereby, we name it the modified-Abe entropy function.
Except for the entropy forms listed in Table~\ref{tab:table1}, there is a well-known entropy---Renyi entropy, which can be related to $S_{h,h'}$ through the relationship between Renyi entropy and Tsallis entropy (only for $q\leq 1$)~\cite{Tsallis2001},
\begin{equation}
    S_q^{\mathrm{Renyi}} \equiv\frac{\ln\sum_{i=1}^{W}p_i^q}{1-q}
    = \frac{\ln\left[1+(1-q)S_q^{\mathrm{Tsallis}}\right]}{1-q}.
\end{equation}

\begin{table}[H]
	\caption{\label{tab:table1}The two-parameter entropy $S_{h,h'}$ recovers other entropy functions by the variation of $h,h'$.}
 \setlength{\tabcolsep}{11.56mm}
	\newcolumntype{C}{>{\centering\arraybackslash}X}
	\begin{tabularx}{\textwidth}{lC}
	\toprule
	\textbf{Entropy Type} & \boldmath{$S_{h,h'}$} \\
	\midrule
	Boltzmann--Gibbs & $h=h'\rightarrow 0$\\
	\midrule
	\multirow{2}[-1]{*}{Tsallis~\cite{Tsallis:1987eu}} & $h=q-1,h'=0$\\
													   & or $h=0,h'=1-q$\\
	\midrule
	$\kappa$~\cite{kaniadakis2001non} & $h'=h=\kappa$\\
	\midrule
	$(\kappa,r)$~\cite{Kaniadakis:2004rj} & $h=r+\kappa, h'=\kappa - r$ \\
	\midrule
	$\gamma$~\cite{Kaniadakis:2004rj} & $h=2\gamma,h'=\gamma$ \\
	\midrule
	\multirow{2}[-1]{*}{Abe~\cite{abe1997note}} & $h=q-1,h'=1-q^{-1}$ \\
										        & or $h=q^{-1}-1,h'=1-q$ \\
	\midrule
	Shafee~\cite{shafee2004generalized,shafee2007lambert,Wang:2003} & $h'\rightarrow -h$ \\
	\midrule
	modified Abe & $h'=-1/h$ \\
	\bottomrule
\end{tabularx}
\end{table}

\section{Properties}

Now we shall address some properties of this $(h,h')$-entropy, $S_{h,h'}$.
As we all know, the Boltzmann--Gibbs and the Tsallis entropy can be expressed as~\cite{tsallis2009,tsallis1994numbers}
\begin{equation}
    S_{BG} = -\left<\ln p_i\right>=\left<\ln\left(1/p_i\right)\right>,\quad S_q = \left<\ln_q\left(1/p_i\right)\right>,
\end{equation}
where $\left<\ldots\right>\equiv\sum_{i=1}^Wp_i\left(\ldots\right)$ is the standard mean value, and $\ln_q$ is $q$-logarithm. Along this line, we straightforwardly obtain
\begin{equation}
    S_{h,h'} = \left<\ln_{h,h'}\left(1/p_i\right)\right>,
\end{equation}
where $\ln_{h,h'}$ is the $(h,h')$-logarithm, and it can be expressed as
\begin{equation}
    \ln_{h,h'}(x) = \frac{x^{h'}-x^{-h}}{h+h'}.
\end{equation}

% {\color{red}Different from the other non-extensive entropic formulas, it is well defined whether or not one or more states have zero probability.
% The parameters $h$ and $h'$, on the other hand, could take some negative
% values, which does not work explicitly for the Tsallis non-extensive parameter,
% $q$, for example.
% Worthy to mention that the corresponding probability distribution functions for Tsallis, Kaniadakis and other non-extensive entropy formalisms with the various limiting values of parameters $h,h'$ have been derived, whereas the probability distribution functions for
% the generalized $h$-derived motivated $S_{h,h'}$ is very difficult to obtain
% mathematically.}

\subsection{Non-Negativity}

First of all, we consider a thermal system within any possible state.
The probability distribution of each microstate $i$ is defined as $p_i$.
If we assume $p_i^{1+h} \geqslant p_i^{1-h'}$, namely, $1+h\leqslant 1-h'$, for $0\leqslant p_i \leqslant 1$, thus can we obtain
$h+h'\leqslant 0$ and this two-parameter entropy $S_{h,h'} \geqslant 0$.

\subsection{Extremal at Equal Probabilities}

Utilizing the Tsallis entropy, $S_q^T = \frac{\summ p_i^q - 1}{1-q}$, this two-parameter entropy $S_{h,h'}$ can be expressed with it as,
\begin{equation}
S_{h,h'} = \frac{1}{h+h'}\left(hS_{1+h}^T + h' S_{1-h'}^T\right).
\end{equation}
\textls[-15]{For the Tsallis entropies inside this formula, namely $S_{1+h}^T$ and $S_{1-h'}^T$, it is easy to know that both of them reach their extreme values when all the probabilities are equal~\cite{tsallis2009}.
Therefore, at the state of equal probability, our entropy $S_{h,h'}$ will also approach to its extreme value since}
\begin{equation}
\frac{{\dd}}{\dd p_i}S_{h,h'}=\frac{1}{h+h'}\left[h\frac{{\dd}}{\dd p_i}S_{1+h}^T+h'\frac{{\dd}}{\dd p_i}S_{1-h'}^T\right]=0.
\end{equation}

\subsection{Expansibility}

It is straightforwardly verified that $S_{h,h'}$ is expansible for any values of $h$ and $h'$, since
\begin{equation}
S_{h,h'}(p_1,p_2,\cdots,p_W,0)=S_{h,h'}(p_1,p_2,\cdots,p_W).
\end{equation}
This property trivially follows from the definition itself.
It means when we add some events with zero probabilities, $S_{h,h'}$ keeps invariant.

\subsection{Non-Additivity}

When we consider a system that can be decomposed into two independent sub-systems, $A$ and $B$, ($p_{ij}^{A+B}=p_i^Ap_j^B$),
\begin{eqnarray}
S_{h,h'}(A+B) &=&
-\sum_{i=1}^{W_A}\sum_{j=1}^{W_B}\frac{\left(p_{ij}^{A+B}\right)^{1+h} -
\left(p_{ij}^{A+B}\right)^{1-h'}}{h+h'} \nonumber\\
&=& S_{h,h'}(A)\cdot \sum_{j=1}^{W_B}\left(p_j^B\right)^{1+h} + S_{h,h'}(B)\cdot
\sum_{i=1}^{W_A}\left(p_i^A\right)^{1-h'}.
\end{eqnarray}
The values of $h$ and $h'$ cannot be zero together in case (or $h=-1$ and $h'=1$ appear at the same time).
In other words, $S_{h,h'}$ is said to be non-additive similar to the Tsallis non-extensive entropy.

% \subsection{About distribution}

% {\bf
% Consider the canonical ensemble, we have the norm constraint Eq.~\eqref{equ:normconstraint} and the energy constraint Eq.~\eqref{equ:energyconstraint}:
% \begin{equation}
% 	\label{equ:normconstraint}
% 	\summ p_i = 1,
% \end{equation}
% \begin{equation}
% 	\label{equ:energyconstraint}
% 	\summ p_iE_i = U.
% \end{equation}
% Use the Lagrange method to find the optimizing distribution, we define
% \begin{equation}
% 	\label{equ:lagrange}
% 	\Phi[p_i] \equiv S_{h,h'} - \alpha \summ p_i - \beta \summ p_iE_i,
% \end{equation}
% where $\alpha$ and $\beta$ are the Lagrange parameters. Then, we impose $\partial\Phi[p_i]/\partial p_i = 0$, and straightforwardly obtain
% \begin{equation}
% 	-(1+h)p_i^h+(1-h')p_i^{h'}=(h+h')(\alpha+\beta E_i).
% \end{equation}
% It is not difficult to find that the above formula can not get a analytical solution.
% }

\section{Summary and Outlook}
To summarize, with the generalized $h$-derivative we firstly propose a two-parameter non-additive entropy, $S_{h,h'}$, in order to connect several non-extensive entropy functions.
The $h$-derivative motivated non-additive entropy, $S_{h,h'}$, is demonstrated to recover different kinds of non-extensive entropy formalisms, such as the Tsallis entropy ($h=q-1, h'=0$), the Abe entropy ($h=q-1, h'=1-q^{-1}$), the Borges--Roditi entropy ($h=q-1, h'=1-q'$), the Kaniadakis $\kappa$-entropy ($h'=h=\kappa$) and the Shafee ($h'\rightarrow -h$ or $h'=-h+\delta$, here $\delta \to 0$) non-extensive entropy by varying values of $h,h'$.
On the other hand, the present two-parameter entropy exhibits all the relative properties as a generalized non-extensive entropy.
Furthermore, the remarkable relationship between $S_{h,h'}$ and other non-extensive entropies may cast a light on the connection of  non-extensive entropy and some mathematical structures such as quantum calculus.
It may lead to a deeper understanding of the mathematical and physical
foundations of non-extensive statistics.
We also noticed some other two-parameter distribution functions, such as the $(r,q)$-distribution and $(\alpha,\kappa)$-distribution~\cite{Qureshi:2014, Abid:2015}, which have been well-applied to astrophysics or space plasma physics. These two-parameter distributions provide another view to investigate the non-Maxwellian systems.
It will be of great interest to associate this $(h,h')$-entropy with them and find out the deeper connections.
There are also various elegant forms of entropy, such as fractional entropy~\cite{UBRIACO20092516} and
Deng entropy~\cite{Deng2020}. Our $(h,h')$-entropy, $S_{h,h'}$, is indeed unable to establish a connection with theirs.
Further exploration of the inherent connections between different forms of entropy is necessary.

\vspace{6pt}

\authorcontributions {Conceptualization, methodology, B.-W.Z.; formal analysis, investigation, J.-W.K. and K.-M.S.;
        visualization, writing---original draft preparation, J.-W.K.; validation, writing---review and editing, K.-M.S.;
        funding acquisition, B.-W.Z. and K.-M.S..
        All authors have read and agreed to the published version of the manuscript.}
    %For research articles with several authors, a short paragraph specifying their individual contributions must be provided. The following statements should be used ``Conceptualization, X.X. and Y.Y.; methodology, X.X.; software, X.X.; validation, X.X., Y.Y. and Z.Z.; formal analysis, X.X.; investigation, X.X.; resources, X.X.; data curation, X.X.; writing---original draft preparation, X.X.; writing---review and editing, X.X.; visualization, X.X.; supervision, X.X.; project administration, X.X.; funding acquisition, Y.Y. All authors have read and agreed to the published version of the manuscript.'', please turn to the  \href{http://img.mdpi.org/data/contributor-role-instruction.pdf}{CRediT taxonomy} for the term explanation. Authorship must be limited to those who have contributed substantially to the work~reported.

\funding{This research was funded by the Guangdong Major Project of Basic and Applied Basic Research No. 2020B030103008, the funding for the Doctoral Research of ECUT (No. DHBK2019211) and Natural Science Foundation of China with Project Nos. 11935007 and 12035007.}

\institutionalreview {Not applicable.}%In this section, you should add the Institutional Review Board Statement and approval number, if relevant to your study. You might choose to exclude this statement if the study did not require ethical approval. Please note that the Editorial Office might ask you for further information. Please add “The study was conducted in accordance with the Declaration of Helsinki, and approved by the Institutional Review Board (or Ethics Committee) of NAME OF INSTITUTE (protocol code XXX and date of approval).” for studies involving humans. OR “The animal study protocol was approved by the Institutional Review Board (or Ethics Committee) of NAME OF INSTITUTE (protocol code XXX and date of approval).” for studies involving animals. OR “Ethical review and approval were waived for this study due to REASON (please provide a detailed justification).” OR “Not applicable” for studies not involving humans or animals.

\dataavailability {Not applicable.}%We encourage all authors of articles published in MDPI journals to share their research data. In this section, please provide details regarding where data supporting reported results can be found, including links to publicly archived datasets analyzed or generated during the study. Where no new data were created, or where data is unavailable due to privacy or ethical re-strictions, a statement is still required. Suggested Data Availability Statements are available in section “MDPI Research Data Policies” at \url{https://www.mdpi.com/ethics}.

\acknowledgments{The author, Ke-Ming Shen, would like to show his grateful thanks for the fruitful discussions with T. S. Biro and C. Y. Yu.}

\conflictsofinterest {The authors declare no conflict of interest.}%Declare conflicts of interest or state ``The authors declare no conflict of interest.'' Authors must identify and declare any personal circumstances or interest that may be perceived as inappropriately influencing the representation or interpretation of reported research results. Any role of the funders in the design of the study; in the collection, analyses or interpretation of data; in the writing of the manuscript; or in the decision to publish the results must be declared in this section. If there is no role, please state ``The funders had no role in the design of the study; in the collection, analyses, or interpretation of data; in the writing of the manuscript; or in the decision to publish the~results''.

%\begin{acknowledgments}
%
%%The research is supported by the Guangdong Major Project of Basic and Applied Basic Research No. 2020B030103008, and Natural Science Foundation of China with Project Nos. 11935007 and 12035007.
%The author, Ke-Ming Shen, would like to show his grateful thanks for the funding for the Doctoral Research of ECUT.
%
%\end{acknowledgments}

%\abbreviations{Abbreviations}{
%	The following abbreviations are used in this manuscript:\\
%
%	\noindent
%	\begin{tabular}{@{}ll}
%		BG & Boltzmann\,--\,Gibbs
%	\end{tabular}
%}

%%%%%%%%%%%%%%%%%%%%%%%%%%%%%%%%%%%%%%%%%%
%% Optional
\appendixtitles{no} % Leave argument "no" if all appendix headings stay EMPTY (then no dot is printed after "Appendix A"). If the appendix sections contain a heading then change the argument to "yes".
\appendixstart
\appendix
%\section[\appendixname~\thesection]{}
\section[\appendixname]{} \label{app1}

%\subsection{Appendix}
S. Abe has proven an interesting property~\cite{abe1997note} that the BG entropy can
be rewritten as a derivative of
\begin{equation}
	S_{BG} = -\frac{\dd}{\dd x}\summ p_i^x\Bigg|_{x=1},
\label{equ:shannonforderivative}
\end{equation}
and the Tsallis one has a similar property
\begin{equation}
	\label{equ:tsallisforqderivative}
	S_q=-D_q\summ p_i^x\Bigg|_{x=1},
\end{equation}
 where $D_q$ is Jackson
 \textit{q}-derivative~\cite{Jackson:1909,Jackson:1910,ernst2000history,aral2013applications},
\begin{equation}
	\label{equ:jacksondefine}
	D_qf(x)\equiv\frac{f(qx)-f(x)}{qx-x}.
\end{equation}
%Obviously, we have $D_1f(x)=\dd f(x)/\dd x$.
%With respect to the symmetry that plays a central role in quantum groups
%\cite{Gomez:1996az},
Abe applied the symmetric $q\leftrightarrow q^{-1}$
to give a new modified \textit{q}-derivative as follows:
\begin{equation}
	\label{equ:abetwoparaq}
	D_{q,q^{-1}}f(x)\equiv\frac{f(qx)-f(q^{-1}x)}{qx-q^{-1}x},
\end{equation}
thus a symmetric modification of Tsallis entropy goes as
\begin{equation}
	\label{equ:abeentropy}
	S_q^S=-\sum_{i=1}^{W}\frac{(p_i)^q-(p_i)^{q^{-1}}}{q-q^{-1}}.
\end{equation}

Inspired by S. Abe, Borges and Roditi define a two-parameter $q$-derivative~\cite{borges1998family}:
\begin{equation}
	\label{equ:twoparaq}
	D_{q,q'}f(x)\equiv\frac{f(qx)-f(q'x)}{qx-q'x}, \quad q,q'\in \mathbb{R},
\end{equation}
and its corresponding entropic form is
\begin{equation}
	\label{equ:borgesentropy}
	S_{q,q'}=-\sum_{i=1}^{W}\frac{p_i^q-p_i^{q'}}{q-q'}.
\end{equation}

\begin{adjustwidth}{-\extralength}{0cm}

\reftitle{References}

\PublishersNote{}
\end{adjustwidth}


\begin{thebibliography}{999}

\bibitem[Dauxois et~al.(2002)Dauxois, Ruffo, Arimondo, and
  Wilkens]{Dauxois2002}
Dauxois, T.; Ruffo, S.; Arimondo, E.; Wilkens, M. Dynamics and Thermodynamics
  of Systems with Long-Range Interactions: An Introduction.
\newblock In {\em Dynamics and Thermodynamics of Systems with Long-Range
  Interactions}; Dauxois, T., Ruffo, S., Arimondo, E., Wilkens, M., Eds.;
  Springer: Berlin/Heidelberg, Germany,
 2002; pp. 1--19.
\newblock {\url{https://doi.org/10.1007/3-540-45835-2_1}}.

\bibitem[Salzberg(1965)]{salzberg1965exact}
Salzberg, A.M.
\newblock Exact statistical thermodynamics of gravitational interactions in one
  and two dimensions.
\newblock {\em J. Math. Phys.} {\bf 1965}, {\em 6},~158--160.

\bibitem[Montroll and Shlesinger(1983)]{montroll1983maximum}
Montroll, E.W.; Shlesinger, M.F.
\newblock Maximum entropy formalism, fractals, scaling phenomena, and 1/f
  noise: A tale of tails.
\newblock {\em J. Stat. Phys.} {\bf 1983}, {\em 32},~209--230.

\bibitem[Tsallis(1995)]{TSALLIS1995539}
Tsallis, C.
\newblock Some comments on Boltzmann-Gibbs statistical mechanics.
\newblock {\em Chaos Solitons Fractals} {\bf 1995}, {\em 6},~539--559.
%\newblock Complex Systems in Computational Physics,
  {\url{https://doi.org/10.1016/0960-0779(95)80062-L}}.

\bibitem[Tsallis(1988)]{Tsallis:1987eu}
Tsallis, C.
\newblock {Possible Generalization of Boltzmann-Gibbs Statistics}.
\newblock {\em J. Stat. Phys.} {\bf 1988}, {\em 52},~479--487.
%\newblock For an updated bibliography on this subject, see
%  \url{http://tsallis.cat.cbpf.br/TEMUCO.pdf},
  {\url{https://doi.org/10.1007/BF01016429}}.

\bibitem[Tsallis(2009)]{tsallis2009}
Tsallis, C.
\newblock {\em Introduction to Nonextensive Statistical Mechanics: Approaching
  a Complex World}; Springer: New York, NY, USA, 2009.
\newblock {\url{https://doi.org/10.1007/978-0-387-85359-8}}.

\bibitem[Plastino and Plastino(1995)]{PLASTINO1995347}
Plastino, A.; Plastino, A.
\newblock Non-extensive statistical mechanics and generalized Fokker-Planck
  equation.
\newblock {\em Phys. A} {\bf 1995}, {\em 222},~347--354.
\newblock {\url{https://doi.org/10.1016/0378-4371(95)00211-1}}.

\bibitem[Bir\'o et~al.(2015)Bir\'o, Shen, and Zhang]{Biro:2014ata}
Bir\'o, T.S.; Shen, K.M.; Zhang, B.W.
\newblock {Non-extensive quantum statistics with particle-hole
  symmetry}.
\newblock {\em Phys. A} {\bf 2015}, {\em 428},~410--415,
%  \href{http://xxx.lanl.gov/abs/1412.2971}{{\normalfont
%  [arXiv:cond-mat.stat-mech/1412.2971]}}.
\newblock {\url{https://doi.org/10.1016/j.physa.2015.01.072}}.

\bibitem[Shen et~al.(2017{\natexlab{a}})Shen, Zhang, and Wang]{Shen:2017qwo}
Shen, K.M.; Zhang, B.W.; Wang, E.K.
\newblock {Generalized Ensemble Theory with Non-extensive Statistics}.
\newblock {\em Phys. A} {\bf 2017}, {\em 487},~215--224,
%  \href{http://xxx.lanl.gov/abs/1707.03526}{{\normalfont
%  [arXiv:cond-mat.stat-mech/1707.03526]}}.
\newblock {\url{https://doi.org/10.1016/j.physa.2017.06.030}}.

\bibitem[Shen et~al.(2017{\natexlab{b}})Shen, Zhang, Hou, Zhang, and
  Wang]{Shen:2017etj}
Shen, K.M.; Zhang, H.; Hou, D.F.; Zhang, B.W.; Wang, E.K.
\newblock {Chiral Phase Transition in Linear Sigma Model with Nonextensive
  Statistical Mechanics}.
\newblock {\em Adv. High Energy Phys.} {\bf 2017}, {\em 2017},~4135329,
%  \href{http://xxx.lanl.gov/abs/1707.02735}{{\normalfont
%  [arXiv:nucl-th/1707.02735]}}.
\newblock {\url{https://doi.org/10.1155/2017/4135329}}.

\bibitem[Khachatryan et~al.(2010{\natexlab{a}})]{CMS:2010tjh}
Khachatryan, V.
\newblock {Transverse-momentum and pseudorapidity distributions of charged
  hadrons in $pp$ collisions at $\sqrt{s}=7$ TeV}.
\newblock {\em Phys. Rev. Lett.} {\bf 2010}, {\em 105},~022002,
%  \href{http://xxx.lanl.gov/abs/1005.3299}{{\normalfont
%  [arXiv:hep-ex/1005.3299]}}.
\newblock {\url{https://doi.org/10.1103/PhysRevLett.105.022002}}.

\bibitem[Khachatryan et~al.(2010{\natexlab{b}})]{CMS:2010wcx}
Khachatryan, V.
\newblock {Transverse Momentum and Pseudorapidity Distributions of Charged
  Hadrons in pp Collisions at $\sqrt{s} = 0.9$ and 2.36 TeV}.
\newblock {\em J. High Energy Phys.} {\bf 2010}, {\em 02},~041,
%  \href{http://xxx.lanl.gov/abs/1002.0621}{{\normalfont
%  [arXiv:hep-ex/1002.0621]}}.
\newblock {\url{https://doi.org/10.1007/JHEP02(2010)041}}.

\bibitem[Betzler and Borges(2012)]{Betzl:2012}
Betzler, A.S.; Borges, E.P.
\newblock Nonextensive distributions of asteroid rotation periods and
  diameters.
\newblock {\em Astron. Astrophys.} {\bf 2012}, {\em 539},~A158.
\newblock {\url{https://doi.org/10.1051/0004-6361/201117767}}.

\bibitem[Wang and Du(2018)]{wang2018viscosity}
Wang, Y.; Du, J.
\newblock The viscosity of charged particles in the weakly ionized plasma with
  power-law distributions.
\newblock {\em Phys. Plasmas} {\bf 2018}, {\em 25},~062309.

\bibitem[Kaniadakis(2001)]{kaniadakis2001non}
Kaniadakis, G.
\newblock Non-linear kinetics underlying generalized statistics.
\newblock {\em Phys. A} {\bf 2001}, {\em 296},~405--425.

\bibitem[Shafee(2004)]{shafee2004generalized}
Shafee, F.
\newblock Generalized Entropies and Statistical Mechanics. \emph{arXiv} {\bf 2004}, arXiv:cond-mat/0409037.
%\newblock  \href{http://xxx.lanl.gov/abs/cond-mat/0409037}{{\normalfont
%  [arXiv:cond-mat.stat-mech/cond-mat/0409037]}}.

\bibitem[Abe(1997)]{abe1997note}
Abe, S.
\newblock A note on the $q$-deformation-theoretic aspect of the generalized
  entropies in nonextensive physics.
\newblock {\em Phys. Lett. A} {\bf 1997}, {\em 224},~326--330.

\bibitem[Borges and Roditi(1998)]{borges1998family}
Borges, E.P.; Roditi, I.
\newblock A family of nonextensive entropies.
\newblock {\em Phys. Lett. A} {\bf 1998}, {\em 246},~399--402.

\bibitem[Kac and Cheung(2002)]{kac2002quantum}
Kac, V.G.; Cheung, P.
\newblock {\em Quantum Calculus}; Springer: Berlin/Heidelberg, Germany,
  2002; Volume 113.

\bibitem[Jagerman(2000)]{jagerman2000difference}
Jagerman, D.L.
\newblock {\em Difference Equations with Applications to Queues}; CRC Press: Boca Raton, FL, USA,
  2000.

\bibitem[Shafee(2007)]{shafee2007lambert}
Shafee, F.
\newblock Lambert function and a new non-extensive form of entropy.
\newblock {\em IMA J. Appl. Math.} {\bf 2007}, {\em
  72},~785--800.

  \bibitem[Tsallis(2001)]{Tsallis2001}
Tsallis, C., I. Nonextensive Statistical Mechanics and Thermodynamics:
  Historical Background and Present Status.
\newblock In {\em Nonextensive Statistical Mechanics and Its Applications};
  Abe, S., Okamoto, Y., Eds.; Springer: Berlin/Heidelberg, Germany,
  2001; pp. 3--98.
\newblock {\url{https://doi.org/10.1007/3-540-40919-X_1}}.

\bibitem[Kaniadakis et~al.(2005)Kaniadakis, Lissia, and
  Scarfone]{Kaniadakis:2004rj}
Kaniadakis, G.; Lissia, M.; Scarfone, A.M.
\newblock {Two-parameter deformations of logarithm, exponential, and entropy: A
  consistent framework for generalized statistical mechanics}.
\newblock {\em Phys. Rev. E} {\bf 2005}, {\em 71},~046128,
%  \href{http://xxx.lanl.gov/abs/cond-mat/0409683}{{\normalfont
%  [cond-mat/0409683]}}.
\newblock {\url{https://doi.org/10.1103/PhysRevE.71.046128}}.

\bibitem[Wang(2003)]{Wang:2003}
Wang, Q.
\newblock Extensive Generalization of Statistical Mechanics Based on Incomplete
  Information Theory.
\newblock {\em Entropy} {\bf 2003}, {\em 5},~220–232.
\newblock {\url{https://doi.org/10.3390/e5020220}}.


\bibitem[Tsallis(1994)]{tsallis1994numbers}
Tsallis, C.
\newblock What are the numbers that experiments provide.
\newblock {\em Quim. Nova} {\bf 1994}, {\em 17},~468--471.

\bibitem[Qureshi et~al.(2014)Qureshi, Nasir, Masood, Yoon, Shah, and
  Schwartz]{Qureshi:2014}
Qureshi, M.N.S.; Nasir, W.; Masood, W.; Yoon, P.H.; Shah, H.A.; Schwartz, S.J.
\newblock Terrestrial lion roars and non-Maxwellian distribution.
\newblock {\em J. Geophys. Res. Space Phys.} {\bf 2014}, {\em
  119},~10059--10067.
\newblock {\url{https://doi.org/https://doi.org/10.1002/2014JA020476}}.

\bibitem[Abid et~al.(2015)Abid, Ali, Du, and Mamun]{Abid:2015}
Abid, A.A.; Ali, S.; Du, J.; Mamun, A.A.
\newblock Vasyliunas-Cairns distribution function for space plasma species.
\newblock {\em Phys. Plasmas} {\bf 2015}, {\em 22},~084507,
%  \href{http://xxx.lanl.gov/abs/https://doi.org/10.1063/1.4928886}{{\normalfont
%  [https://doi.org/10.1063/1.4928886]}}.
\newblock {\url{https://doi.org/10.1063/1.4928886}}.

\bibitem[Ubriaco(2009)]{UBRIACO20092516}
Ubriaco, M.R.
\newblock Entropies based on fractional calculus.
\newblock {\em Phys. Lett. A} {\bf 2009}, {\em 373},~2516--2519.
\newblock
  {\url{https://doi.org/https://doi.org/10.1016/j.physleta.2009.05.026}}.

\bibitem[Deng(2020)]{Deng2020}
Deng, Y.
\newblock Uncertainty measure in evidence theory.
\newblock {\em Sci. China Inf. Sci.} {\bf 2020}, {\em
  63},~210201.
\newblock {\url{https://doi.org/10.1007/s11432-020-3006-9}}.

\bibitem[Jackson(1909)]{Jackson:1909}
Jackson, F.H. Generalization of the differential operative symbol with an extended form of Boole's equation.
\newblock {\em Mess. Math.} {\bf 1909}, {\em 38},~57.

\bibitem[Jackson(1910)]{Jackson:1910}
Jackson, F.H. On $q$-definite integrals.
\newblock {\em Quart. J. Pure Appl. Math.} {\bf 1910}, {\em 41},~193.

\bibitem[Ernst(2000)]{ernst2000history}
Ernst, T.
\newblock {\em The History of Q-Calculus and a New Method}; UUDM Report;
  Department of Mathematics, Uppsala University: Uppsala, Sweden, 2000.

\bibitem[Aral et~al.(2013)Aral, Gupta, and Agarwal]{aral2013applications}
Aral, A.; Gupta, V.; Agarwal, R.
\newblock {\em Applications of q-Calculus in Operator Theory}; SpringerLink:
  B{\"u}cher; Springer: New York, NY, USA,  2013.

\end{thebibliography}
\end{document}